\documentclass[onecollarge,natbib]{svjour2}
\bibpunct{[}{]}{;}{n}{}{,} 
\smartqed  
\usepackage{graphicx}
\usepackage{bbm}
\newcommand {\ham}{{\mathcal H}}
\newcommand{\bm}[1]{ \mbox{\boldmath $#1$}  }

\journalname{Archive of Applied Mechanics}
\begin{document}

\title{Three-body continuum wave functions with a box boundary condition\thanks{This work was 
supported by funds provided by DGI of MINECO (Spain) under contract No. FIS2011-23565.  }
}


\author{E. Garrido}

\institute{E. Garrido \at
              Instituto de Estructura de la Materia, CSIC, Serrano 123, E-28042 Madrid, Spain \\
              \email{e.garrido@csic.es}           
              }

\date{Received: date / Accepted: date}

\maketitle

\begin{abstract}
In this work we investigate the connection between discretized three-body continuum wave functions, in
particular via a box boundary condition, and the wave functions computed with the correct asymptotics. 
The three-body wave functions are in both cases obtained by means of the adiabatic expansion method.
The information concerning all the possible incoming and outgoing channels, which appears naturally 
when the continuum is not discretized, seems to be lost when the discretization is implemented. In 
this work we show that both methods are fully equivalent, and the full information contained in the
three-body wave function is actually preserved in the discrete spectrum. Therefore, in those cases
when the asymptotic behaviour is not known analytically, i.e., when the Coulomb interaction is involved,
the discretization technique can be safely used.

\end{abstract}

\section{Introduction}
\label{intro}

The description of an $N$-body system in the continuum requires knowledge of three different ingredients: 
How the $N$ particles in the system get close to each other, how they interact, and how the particles move 
far apart from each other. In other words, the description of a system in the continuum is equivalent 
to describing the collision between the particles involved in the system under investigation. 

For two-body systems the problem is relatively simple. Let us consider two point-like particles
approaching each other with a relative energy $E$ and some relative angular momentum 
$\ell$. If the the two-body Hamiltonian ${\cal H}$ commutes with the $\hat{\ell}^2$ operator,
we then have that $\ell$ is 
a good quantum number, and the only possible outgoing channel is the one with the two particles moving far 
apart from each other with the same relative energy, and the same relative angular momentum. 
In this simple single-channel case the continuum two-body wave function is fully determined by one 
characteristic number, the phase shift.  Let us now go a step beyond, and assume that 
$[{\cal H},\hat{\ell}^2] \neq 0$, which implies that the two-body potential couples different 
orbital angular momenta. In this case, when the two particles approach each 
other with relative energy $E$ and angular momentum $\ell$, several outgoing channels are possible, 
namely, all those with relative momentum $\ell^\prime$ such that $\ell$ and $\ell^\prime$ are coupled 
by the potential. Furthermore, in this multichannel case, full knowledge of all possible 
two-body processes requires knowledge of all the outgoing
channels for all the possible incoming channels. In other words, in this case, the two-body reactions are
fully determined, not by a number, but by a matrix (the ${\cal S}$-matrix),
whose $ij$-term, ${\cal S}_{ij}$, is such that $|{\cal S}_{ij}|^2$ gives the probability for an incoming 
channel $i$ to go out through channel $j$. Since given an incoming channel $i$ the system
must necessarily go out through one of the available outgoing channels, it is trivial to conclude
that the following condition, $\sum_j |{\cal S}_{ij}|^2=1$, has to be fulfilled for each $i$. This condition
can be written in a more compact way as the unitarity condition ${\cal S}^\dag {\cal S}=\mathbbm{I}$,
where $\mathbbm{I}$ is the unity matrix.

For a three-particle system the continuum problem is equivalent to the multichannel problem described above.
The total three-body energy $E$ is the sum of the relative energy between two of the particles and
the relative energy between the third particle and the center of mass of the other two. These two 
relative energies are not fixed, and therefore there will be infinitely many channels consistent
with one given total three-body energy. The same happens with the total angular momentum, which in 
general can be obtained by coupling different partial-wave components in the three-body system.
Usually the partial-wave components are coupled by the interaction between the three particles.
Furthermore, if some of the internal two-body subsystems can hold one or more two-body bound states, then
additional available three-body channels appear. These are usually denoted as 1+2 channels, and
they correspond to a bound two-body system and the third particle in the continuum.  In general, 
the interactions between the particles are able to couple the different channels, in such a way that,
for instance, three incoming particles in the continuum could go out, when
available, in a 1+2 channel (recombination 
process) or go out keeping the three particles in the continuum (where infinitely many distributions
of the total three-body energy is in principle possible). In the same way, if we consider an incoming
1+2 channel, provided that the incident energy is big enough, the system could go out through
the same or another 1+2 channel (giving rise to elastic, inelastic, or transfer reactions) or lead to
the three particles in the continuum (breakup process). The full information about all these possible 
reactions is contained in the three-body ${\cal S}$ matrix, which connects all the possible incoming
channels with all the possible outgoing channels.

\begin{figure}
\centering
\includegraphics[angle=-90,scale=0.3]{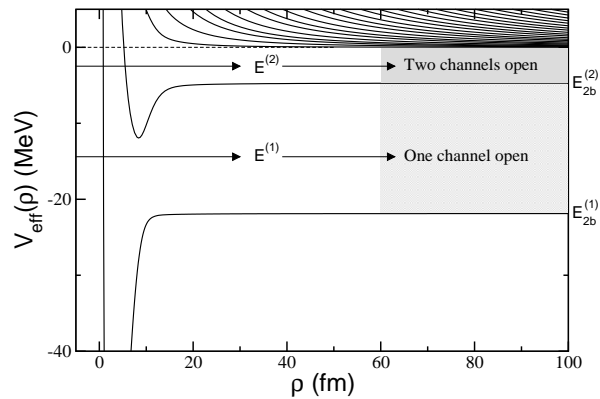}
\caption{ Typical effective adiabatic potentials for a three-body system where two two-body bound
states are present. The two lowest adiabatic
potentials go asymptotically to the binding energies $E_{2b}^{(1)}$ and $E_{2b}^{(2)}$ of the
two-body bound states. For a given three-body energy $E$, when $E_{2b}^{(1)}<E<E_{2b}^{(2)}$ only
one channel is open, while when $E_{2b}^{(2)}<E<0$ both channels are open.}
\label{fig1}    
\end{figure}

The first difficulty when describing three-body continuum states is the identification of the
different incoming and outgoing channels. In this connection the use of the adiabatic expansion 
method \cite{nie01} is particularly useful \cite{mac02}. In this method the three-body problem is reduced to
a coupled set of radial equations where a family of effective adiabatic potentials enter. The
wave function is expanded in an adiabatic basis such that each term in the expansion is
associated to a single adiabatic potential. Therefore, in this basis the dimension of the ${\cal S}$-matrix
describing the full process is determined by the number of adiabatic terms included in the
expansion, which is typically a rather modest number. The great advantage of the adiabatic expansion
is that, asymptotically, the possible 1+2 channels and the breakup channels (the three-particles
in the continuum) are clearly separated.
As an illustration, we show in Fig.\ref{fig1} a typical set of adiabatic potentials entering in 
a three-body problem.  They correspond
to a three-body system where two of the two-body subsystems have a bound state. This is reflected
in the fact that the two lowest effective adiabatic potentials go asymptotically to the binding
energies $E_{2b}^{(1)}$ and $E_{2b}^{(2)}$ of each bound two-body system.
As shown for instance in \cite{rom11}, the channels associated to these adiabatic terms 
correspond to asymptotic structures where two of the particles form a bound state (with
energy $E_{2b}^{(1)}$ or $E_{2b}^{(2)}$) and the third particle moves in the continuum.
In other words, these channels describe the 1+2 incoming or outgoing channels, which 
are separated from the infinitely many remaining ones, which describe the three particles
in the continuum.

In this way we can distinguish different regions in Fig.\ref{fig1} depending on the total 
three-body energy.  All the three-body energies $E$ such that
$E_{2b}^{(1)} < E < E_{2b}^{(2)}$ (like $E^{(1)}$ in the figure) correspond to processes where
only one channel is open. Only the elastic collision between the third particle
and the bound two-body state with energy $E_{2b}^{(1)}$ is possible. When the three-body energy
increases up to the region $E_{2b}^{(2)} < E < 0$ ($E^{(2)}$ in the figure) a second channel is
open. Two different collisions are now possible, the one where a particle hits the bound state
with binding energy $E_{2b}^{(1)}$, and the one where a particle hits the state with binding
energy $E_{2b}^{(2)}$. In the same way, each of these reactions has two possible outgoing channels, 
corresponding to the two allowed bound two-body states and the third particle in the continuum. 
In particular, in this energy range the rearrangement process is open. The full process will be then
described by just a $2\times2$ ${\cal S}$-matrix. When $E>0$ the breakup channels are also open, and
in principle infinitely many channels should be included in order to describe the reaction. 

The second problem to be faced when describing three particles in the continuum is the 
normalization of the three-body wave function. When long-range interactions are not involved
in the problem, the asymptotic behaviour is known analytically, 
and the continuum wave function can be normalized to match the expected asymptotic form.
The ${\cal S}$ matrix can be extracted for each incoming and outgoing channel by direct
comparison with the asymptotics, or more accurately, by the use of the integral relations
recently derived \cite{rom11,bar09}.
However, when at least two of the three particles are charged, the asymptotic form of the
wave function is not known, and its normalization becomes a complicated task. One option to solve
this problem is to obtain numerically the order and the Sommerfeld parameter of the regular 
and irregular Coulomb functions that determine the asymptotic outgoing wave for each of the
open channels. The problem is that an inaccurate calculation of these parameters unavoidably
leads to a wrong value of the computed ${\cal S}$-matrix.

In this work we shall consider a different method in order to achieve proper asymptotic behaviour.
This is the discretization of the continuum states. One of the simplest procedures to do so is to
impose a box boundary condition. This procedure immediately leads to a set of discrete continuum states which
are formally treated as bound states, and therefore they are just normalized to 1 inside the box.
Playing with the size of the box it is possible to obtain the discrete continuum state for a 
particular value of the three-body energy. The apparent inconsistency of this method is that 
the discretization inside a box automatically leads to a set discrete energies, each of them
corresponding to a single three-body wave function. No condition is given in order to determine
the incoming and outgoing channels for each of the states, and therefore it is not clear to what 
incoming and outgoing channels each discretized state corresponds. In fact, for a given energy one 
should get
as many different three-body wave functions as channels are open, and not only one. Therefore, 
apparently, a large part of the information has been lost when discretizing the spectrum.
The purpose of this work is to show that this is not correct. The discrete 
continuum states keep the full information about the three-body state. All the possible
incoming and outgoing channels are actually taken into account, and therefore the 
information contained in the ${\cal S}$-matrix is fully preserved. 

In the next two sections we briefly describe the adiabatic expansion method and the
two different procedures used to compute the three-body wave functions. 
They are illustrated by means of a three-body system made of three identical spinless
bosons coupled to spin and parity $0^+$. In the fourth section we discuss the details of
the discretized spectrum, and show that it is equivalent to the results obtained without
discretization. We close the paper with the summary and the conclusions.

%

\section{Three-body continuum wave functions and the adiabatic expansion method.}

The three-body wave functions will be written in terms of the usual $\bm{x}$ and $\bm{y}$ Jacobi coordinates. 
From them we can construct the hyperspherical coordinates \cite{nie01}, which contain a radial one, the 
so-called hyperradius $\rho$ ($\rho^2=\sqrt{x^2+y^2}$) and the five hyperangles  
$\Omega$, i.e, $\alpha=\arctan(x/y)$, $\Omega_x$, and $\Omega_y$.

In hyperspherical coordinates the Hamiltonian operator $\hat{\ham}$ takes the form:
\begin{equation}
\hat{\ham} =  
 -\frac{\hbar^2}{2 m} \hat{T}_\rho + \hat{{\cal H}}_\Omega    ,
\label{eq1}
\end{equation}
where $\hat{T}_\rho$ 
is the hyperradial kinetic energy operator, and $\hat{{\cal H}}_\Omega$ contains the whole
dependence on the hyperangles. 

In the adiabatic expansion method the angular part of the Hamiltonian is first solved for
fixed values of the hyperradius.  This amounts to solve the eigenvalue problem 
\begin{equation}
\hat{{\cal H}}_\Omega \Phi_n(\rho,\Omega)=\frac{\hbar^2}{2 m} \frac{1}{\rho^2}
\lambda_n(\rho) \Phi_n(\rho,\Omega)
\label{eq2}
\end{equation}
for individual values of $\rho$, which is treated as a parameter. 

The angular functions $\{\Phi_n(\rho, \Omega)\}$ form an orthonormal basis for each value of $\rho$,
and they are used to expand the total three-body wave function as:
\begin{equation}
\Psi_i(\bm{x},\bm{y}) = \frac{1}{\rho^{5/2}}\sum_{n=1}^\infty f_{ni}(\rho) \Phi_n(\rho,\Omega),
\label{eq3}
\end{equation}
where $i=1,\cdots, n_0$ labels the incoming channel and $n_0$ is the number of open channels.
Obviously the summation above has to be truncated, and only a finite number $n_A$ of adiabatic terms will
be in practice included in an actual calculation. Note that $n_0$ and $n_A$ are not necessarily equal.
As an example, for the energy $E^{(2)}$ shown in Fig.\ref{fig1} we have that $n_0=2$, and $n_A$ will be 
as large a needed in order to get convergence in whatever observable is computed with the wave 
function (\ref{eq3}).

In a second step, the radial wave functions $f_{ni}(\rho)$ in the expansion in Eq.(\ref{eq3}) are obtained
after solving the following coupled set of radial equations:
\begin{equation}
\sum_{n'=1}^{n_A}\left(\hat{\cal H}_{nn'}-E \delta_{nn'}\right) f_{n'i}(\rho)=0,
\hspace*{1cm} (n^\prime=1,2,\cdots,n_A)
\label{eq4}
\end{equation}
where the operator $\hat{\cal H}_{nn'}$ acts on the radial functions and takes the form
\begin{equation}
\hat{\cal H}_{nn}(\rho)=\frac{\hbar^2}{2m}
\left[ -\frac{d^2}{d\rho^2} -Q_{nn}(\rho) + \frac{1}{\rho^2}
\left( \lambda_n(\rho)+\frac{15}{4} \right) \right]
\label{eq5}
\end{equation}
\begin{equation}
\hat{\cal H}_{nn'}= -\frac{\hbar^2}{2m}
\left( 2 P_{n n'}(\rho) \frac{d}{d\rho} + Q_{n n'}(\rho) \right)
\label{eq6}
\end{equation}
for $n=n'$ and $n\neq n'$, respectively. The coupling terms $P_{nn'}$ and $Q_{nn'}$ are given for instance in Ref.\cite{nie01}. 

The set of coupled differential equations given in Eq.(\ref{eq4}) can be written in a matrix form as:
\begin{small}
\begin{equation}
\left(
        \begin{array}{ccc}
        \hat{\cal H}_{11}-E &  \cdots & \hat{\cal H}_{1n_A} \\
        \hat{\cal H}_{21} &  \cdots & \hat{\cal H}_{2n_A} \\
         \vdots       &  \vdots & \vdots        \\
        \hat{\cal H}_{n_A1} & \cdots & \hat{\cal H}_{n_An_A}-E
        \end{array}
\right)
\left(
        \begin{array}{cccc}
        f_{11}  & f_{12}  & \cdots   & f_{1n_0}\\
        f_{21}  & f_{22}  & \cdots   & f_{2n_0}\\
         \vdots & \vdots  & \vdots   & \vdots \\
        f_{n_A1}& f_{n_A2}& \cdots   & f_{n_An_0}
        \end{array}
\right)
=0,
\end{equation}
\end{small}
and the full three-body wave function is written as:
\begin{small}
\begin{equation}
\Psi=
\left(
        \begin{array}{c}
        \Psi_1  \\
        \Psi_2  \\
         \vdots        \\
        \Psi_{n_0}
        \end{array}
\right)
= \frac{1}{\rho^{5/2}}
\left(
        \begin{array}{cccc}
        f_{11}  & f_{21}  & \cdots   & f_{n_A1}\\
        f_{12}  & f_{22}  & \cdots   & f_{n_A2}\\
         \vdots & \vdots  & \vdots   & \vdots \\
        f_{1n_0}& f_{2n_0}& \cdots   & f_{n_An_0}
        \end{array}
\right)
\left(
        \begin{array}{c}
        \Phi_1  \\
        \Phi_2  \\
         \vdots        \\
        \Phi_{n_A}
        \end{array}
\right),
\label{eq8}
\end{equation}
\end{small}
which contains all the possible incoming channels.

It is important to note that the diagonal operator $\hat{\cal H}_{nn}$ in Eq.(\ref{eq5}) contains the
angular eigenvalues $\lambda_n(\rho)$ introduced in Eq.(\ref{eq2}). They appear in the effective 
adiabatic potentials, which are given by:
\begin{equation}
V^{(n)}_{eff}(\rho)=\frac{\hbar^2}{2m}\left( \frac{\lambda_n(\rho)+\frac{15}{4}}{\rho^2}-Q_{nn}(\rho) \right),
\label{eq9}
\end{equation}
and whose typical behaviour is shown in Fig.\ref{fig1}.

\begin{figure}
\centering
\includegraphics[scale=0.6]{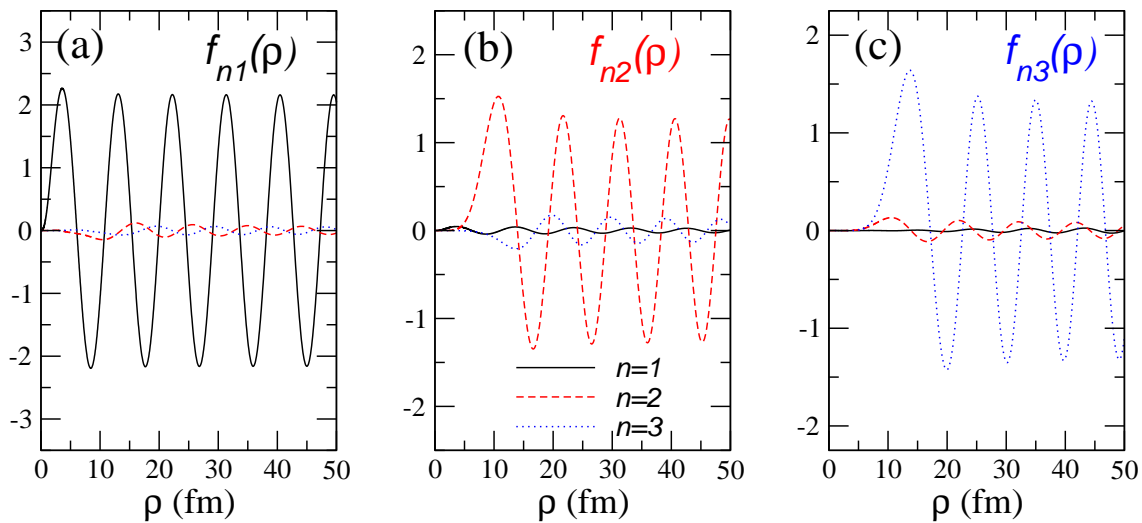}
\caption{Continuum radial wave functions $f_{ni}(\rho)$ contained in Eq.(\ref{eq3}) for incoming channels
 $i=1$ (panel a), $i=2$ (panel b), and $i=3$ (panel c). 
The calculation corresponds to three spinless identical bosons with total spin and parity $0^+$. 
An arbitrary weakly attractive short-range potential has been used. The curves in the figure correspond to
a total three-body energy of $E=10$ MeV.
The calculation has been done including 3 adiabatic
terms in the expansion, and the corresponding functions for $n=1$ are given by the solid curves, $n=2$ by the
dashed curves, and $n=3$ by the dotted curves, respectively. }
\label{fig2}       
\end{figure}

For short-range interactions the asymptotic behaviour of the wave function $\Psi$ is known to be \cite{gar12}:
\begin{equation}
\Psi \rightarrow F - {\cal K} G,
\label{asy}
\end{equation}
where ${\cal K}$ is the ${\cal K}$-matrix, which is related to the ${\cal S}$-matrix by
the simple expression ${\cal S}=(1+i{\cal K})(1-i{\cal K})^{-1}$. In the equation above
$F$ and $G$ are column vectors whose $n^{th}$ term is given by \cite{rom11,gar12}
\begin{equation}
F_n=\sqrt{\kappa} j_{K+ \frac{3}{2}}(\kappa\rho) \frac{1}{\rho^{3/2}} \Phi_n(\rho,\Omega),
\label{eq11}
\end{equation}
\begin{equation}
G_n=\sqrt{\kappa} \eta_{K+ \frac{3}{2}}(\kappa\rho) \frac{1}{\rho^{3/2}} \Phi_n(\rho,\Omega),
\label{eq12}
\end{equation}
where $n=1,\cdots,n_0$, $j$ and $\eta$ are the regular and irregular spherical Bessel functions,
respectively, $K$ is the hypermomentum associated asymptotically to the adiabatic potential $n$,
and $\kappa=\sqrt{2mE}/\hbar$ is the three-body momentum. When $n$ refers to a 1+2 channel 
Eqs.(\ref{eq11}) and (\ref{eq12}) have to be modified as given in Eqs.(12) and (13) in Ref.\cite{gar12}.
This asymptotic behaviour determines how the wave function has to be normalized
in order to extract from it the $n_0 \times n_0$ ${\cal K}$-matrix (and therefore the
${\cal S}$-matrix). When the Coulomb interaction is
involved, the asymptotic behaviour of the wave function is still given by (\ref{asy}),
but the Bessel functions contained in the vectors $F$ and $G$ have to be replaced by
regular and irregular Coulomb functions whose order and Sommerfeld parameter have to be determined numerically.

As an example, we have considered a simple three-body system made of three identical spinless bosons
with total spin and parity $0^+$. We have chosen an arbitrary simple weakly attractive short-range
potential between the three particles. The three-body wave function in Eq.(\ref{eq8}) has then been 
computed and normalized as given in Eq.(\ref{asy}). The corresponding radial wave functions $f_{ni}(\rho)$ are
shown in Fig.\ref{fig2} for a three-body energy of $E=10$ MeV. The calculation has been performed 
including three adiabatic terms in the expansion, and for the energy chosen all the channels are open 
(therefore $n_0=n_A=3$ in Eq.(\ref{eq8})). In the figure we show the radial wave functions 
when the incoming channels are the first adiabatic term ($i=1$, panel a), the second adiabatic
term ($i=2$, panel b), and the third adiabatic term ($i=3$, panel c). The functions corresponding
to outgoing channels $n=$1, 2, and 3, are shown by the solid, dashed, and dotted curves, respectively.
Therefore, Fig.\ref{fig2} shows three different continuum wave function for the same system and the same
total energy.  As we can see in the figure, for each incoming channel $i$ the dominant term is the one with $n=i$.

\section{Discrete continuum states after imposing a box boundary condition.}

As mentioned in the introduction, the continuum spectrum can be easily discretized by imposing a
box boundary condition. This amounts to solving the coupled set of radial equations given in Eq.(\ref{eq4})
imposing the radial wave functions to be zero at a given value $L$ of the hyperradius. This condition
is satisfied only by a discrete set of values of the three-body energy $\{E_i\}$, each of them
associated to a three-body state whose wave function can be written formally as in Eq.(\ref{eq3}): 
\begin{equation}
\Psi^{(i)}(\bm{x},\bm{y}) = \frac{1}{\rho^{5/2}}\sum_{n=1}^\infty f_{n}^{(i)}(\rho) \Phi_n(\rho,\Omega),
\label{eq13}
\end{equation}
but with the important difference that the index $i$ in the expression above labels the wave function
of the three-body state with discrete energy $E_i$. In Eq.(\ref{eq3}) the index $i$ labels the incoming
channel. 

\begin{figure}
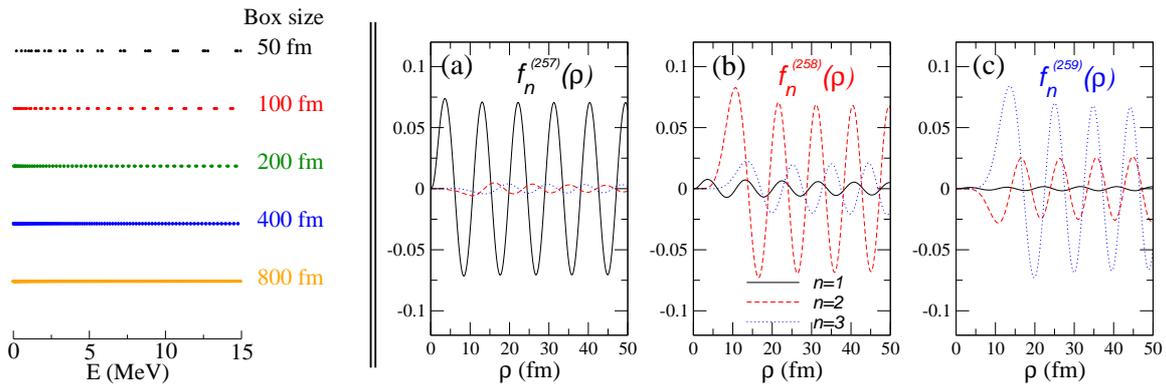

\centering
\includegraphics[scale=0.26]{fig3.eps}
\includegraphics[scale=0.4]{fig4.eps}
\caption{Left: Discrete continuum states for the same system as in Fig.\ref{fig2} for five different
sizes of the normalization box. Right: Continuum radial wave functions $f_n^{(i)}(\rho)$ contained in 
Eq.(\ref{eq13}) for the discrete states $i=257$ with energy 9.99 MeV (panel a), $i=258$ with
energy 10.02 MeV(panel b), and $i=259$ with energy 10.03 MeV (panel c) obtained with a discretization box 
with $L=400$ fm. The calculation has been done including 3 adiabatic terms in the expansion, and the 
corresponding functions for $n=1$, $n=2$, and $n=3$  are given by the solid, dashed, and dotted curves,
respectively. }
\label{fig3} 
\end{figure}

After discretization, it is simple to see that the energy separation between two consecutive discrete
states is given by \cite{gar14}:
\begin{equation}
\Delta E= \frac{2\pi}{L} \frac{E_i}{\kappa_i},
\label{eq14}
\end{equation}
where $\kappa_i=\sqrt{2mE_i}/\hbar$, and where we have assumed that $i$ is large enough.
It is then clear that the larger the size of the box the smaller the separation between the
discrete states, which decreases linearly with $L$. For the same reason, the number of discrete states 
with energy smaller than a given value increases linearly with $L$. This is illustrated in the left
part of Fig.\ref{fig3}, where we show the discrete energies smaller than 15 MeV for the same three-body 
system as in Fig.\ref{fig2}. The result for box sizes of $L=$50 fm, 100 fm, 200 fm, 400 fm, and 800 fm 
(from the top to the bottom in the figure) are shown. We can clearly see that, at the scale of the figure,
the discrete energies for $L=$ 400 fm and 800 fm can hardly be distinguished from each other. The number
of discrete states with energies smaller than 15 MeV are 33, 72, 153, 315, and 642 when we move from the 
smallest to the biggest box, which follow very closely the expected linear dependence on $L$.

\section{A closer look into the discrete continuum states.}

The interesting thing about the discrete energies shown in the left part of Fig.\ref{fig3} is that they
are distributed in a non-uniform way. They actually appear in groups (or bins) of states with
rather similar energies. The number of states in each bin is equal to the number of adiabatic channels
used in the expansion in Eq.(\ref{eq3}), which has been taken equal to 3 in the calculation shown in the 
figure. This grouping of 
the states appears no matter the value of $L$ chosen, as one could see by making a zoom of the figure 
for the cases with large $L$.

As also seen in the left part of Fig.\ref{fig3}, the bigger the box the closer the states in the same
bin to degeneracy. In fact, the rule for the energy separation between states given in 
Eq.(\ref{eq14}) is satisfied not by the individual states, but by the bins themselves. As an example,
for a box of 400 fm and $E_i=80$ MeV, the expected energy separation according to Eq.(\ref{eq14})
is of $\sim 0.6$ MeV. When doing the discretization, the energy bins $\{79.17,79.23,79.27\}$ MeV
and $\{79.81,79.87,79.89\}$ MeV are found.

From the discussion above, it is then clear that, first, for an infinitely big box, the bin structure
has disappeared, in such a way that we have, for each energy, as many wave functions as possible incoming
channels. And second, for a finite box the expected
energy separation between the discrete states corresponds to the energy separation between the bins. 
From these facts, one can almost unavoidably conclude that the discrete states belonging to
the same bin do actually describe the full wave function, as given in Eq.(\ref{eq8}), at the energy 
of the bin (whose energy can be approximated by an average of the energies of all the states in the bin). 
In other words, the wave function contains all the terms associated to all the possible incoming channels. 
Also, it is important to keep in mind that the discrete 
continuum states are normalized to 1 inside the discretization box. Therefore, for sufficient large
values of $\rho$ the continuum wave function will not in general behave as given in Eq.(\ref{asy}),
but instead it will go as:
\begin{equation}
\Psi \rightarrow A(F - {\cal K} G),
\label{asy2}
\end{equation}
where $A$ is a normalization constant $n_0\times n_0$ matrix, which can be
computed as $A=F_b\cdot F_c^{-1}$, where $F_b$ and $F_c$ are, respectively, 
the matrices containing the radial functions, see Eq.(\ref{eq8}), obtained with and without 
discretization of the continuum. This matrix is found to be independent of $\rho$.

\begin{figure}
\centering
\includegraphics[scale=0.6]{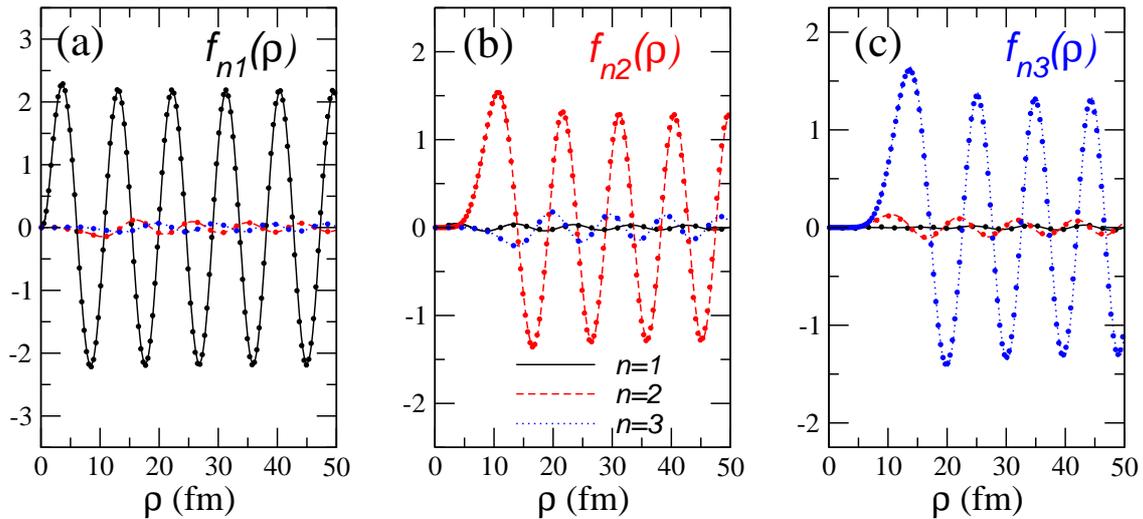}
\caption{Same as in Fig.\ref{fig2} where we have put on top the solid circles that show the discrete
continuum wave functions given in Fig.\ref{fig3} after making the transformation 
$\Psi \rightarrow A^{-1}\Psi$, where the matrix $A$ is defined according to Eq.(\ref{asy2}).}
\label{fig5}
\end{figure}

The conclusion discussed above is confirmed when investigating the radial wave functions. Let us consider 
the same three-body system as the one used in Fig.\ref{fig2}, whose discrete energies have been  shown in 
the left part of Fig.\ref{fig3}. In the right part of the figure, we show in panels (a), (b), and (c) 
the radial wave 
functions contained in Eq.(\ref{eq13}) for the discrete states with labels $i=257$, $i=258$, and $i=259$, 
obtained with a discretization box with $L=400$ fm. The energies of these states are 9.99 MeV, 10.02 MeV, 
and 10.03 MeV, respectively, which are very similar to the one used in the calculation shown in 
Fig.\ref{fig2} (10 MeV). As we can see, the dominant term 
on each panel are the $n=1$, $n=2$, and $n=3$ terms, respectively, exactly as in Fig.\ref{fig2}. In fact, 
a quick eye inspection reveals that the shape of the dominant term on each panel is very similar 
to the corresponding curve in Fig.\ref{fig2}. However, the same quick eye inspection also reveals that the 
non-dominant terms are different (this is particularly clear in panels b and c). Nevertheless, this fact 
does not imply that the wave functions in Figs.\ref{fig2} and \ref{fig3} are not consistent with each other. 
The reason is that 
the wave functions shown in Fig.\ref{fig2} are normalized according to Eq.(\ref{asy}), and the ones in 
Fig.\ref{fig3} are normalized according Eq.(\ref{asy2}). This means that a correct comparison between the 
two wave functions requires to multiply $\Psi$ in Eq.(\ref{asy2}) from the left by $A^{-1}$ (or
$\Psi$ in Eq.(\ref{asy}) from the left by $A$).  When this is done we can really see that the wave functions
shown in Fig.\ref{fig2} and the discretized wave functions shown in Fig.\ref{fig3} do really describe
the same continuum states. This is shown in Fig.\ref{fig5}, where we show the same radial wave functions
as in Fig.\ref{fig2} (thin curves), and where the solid circles  are the corresponding discretized functions 
shown in the right part of Fig.\ref{fig3} but after multiplying $\Psi$ from the left by $A^{-1}$.
The perfect agreement between the thin curves and the solid circles shows that after discretization of the
continuum all the possible incoming and outgoing channels are still considered, and no information is 
actually lost in the process.

\section{Summary and conclusions.}

In this work we have described two different methods that can be used to compute continuum wave functions, namely,
a full continuum treatment normalizing the wave functions according to Eq.(\ref{asy}), or a discretization of
the continuum imposing a box boundary condition and normalizing the discrete wave functions to 1 inside the box.
In particular, we have considered three-body systems and the wave function has been obtained using the 
adiabatic expansion method.

When using the first method, it is rather simple to identify the possible incoming and outgoing channels, 
in such a way that for a given total energy $E$ one can construct the three-body wave function for each of 
the possible incoming channels, and from it the ${\cal K}$- (or ${\cal S}$-) matrix can be obtained. 
However, when using the second method, just a family of discrete energies and wave functions appear. In principle,
for each energy there is only one wave function associated. Since, according to the formal theory, we should
have for each energy as many wave functions as open channels, one could easily conclude that in the process
of discretization part of the information contained in the total wave function has been lost.

In this work we have shown that this is actually not true. When discretizing the spectrum the discrete states
appear organized in bins of states. Each bin has as many elements as open channels are involved in the
calculation. Furthermore, the bigger the discretization box the closer the states in each bin to degeneracy.
Eventually, for an infinitely big box, all the discrete states in a bin would appear at precisely the same 
energy.
We have seen that, in fact, the wave functions of a bin of discrete states perfectly agree with the 
ones corresponding to different incoming channels in a non-discretized calculation.

Therefore, after discretizing the spectrum by imposing a box boundary condition the full information 
about the 
continuum wave function is preserved. This is particularly important when dealing with systems where the 
Coulomb interaction is involved (as the three-alpha system, for instance). In this case the asymptotic 
behaviour of the wave function is not known analytically, and therefore the correct normalization of the
continuum wave function as given in Eq.(\ref{asy}) can become a quite difficult task. However, the 
discretization 
procedure does not require knowledge of the asymptotic form of the wave function, and the treatment of
a system where the Coulomb interaction enters is formally identical to the one of a system where only
short-range interactions are involved.




\end{document}